\documentclass[preprint,12pt]{elsarticle}

\usepackage{amssymb}
\usepackage{amsmath}
\usepackage{mathrsfs}
\usepackage{subfig}
\usepackage{tabularx}
\usepackage{booktabs}
\usepackage{fancyhdr}
\usepackage{cases}
\usepackage{galois}

\begin{document}

\begin{frontmatter}



\title{Quantum public-key encryption schemes based on conjugate coding}


\author{Li Yang\footnote{Corresponding author. E-mail: yangli@iie.ac.cn}, Biyao Yang, Chong Xiang}

\address{State Key Laboratory of Information Security, Institute of Information Engineering, Chinese Academy of Sciences, Beijing 100093, China}

\newtheorem{theorem}{Theorem}
\newtheorem{lemma}[theorem]{Lemma}
\newtheorem{conjecture}[theorem]{Conjecture}
\newtheorem{corollary}[theorem]{Corollary}
\newtheorem{definition}{Definition}
\newtheorem{proposition}[theorem]{Proposition}
\newtheorem{property}{Property}
\newtheorem{scheme}{Scheme~}

\begin{abstract}
We present several quantum public-key encryption (QPKE)
protocols designed with conjugate coding single-photon
string, thus may be realized in laboratory with nowadays
techniques. Two of these schemes are orienting one-bit message, and are extended to two kinds
of QPKE schemes orienting multi-bits. The novel structure of these
protocols ensures they are information-theoretically secure with, probably, a bound greater than any given polynomial of $n$.
Finally, we describe a way to conceal the classical part of the public key with quantum state, this idea is expected to enhance a scheme
to be information-theoretically secure.
\end{abstract}


\end{frontmatter}


\section{Introduction}
\label{} Public-key encryption was first proposed in 1976
\cite{Rivest78,ElGamal85}, which makes the two parties can do
secret communication without presharing a secret key. Up till now,
the security of a public-key cryptography scheme is based on
a mathematically difficult problem, though whose difficulty has not
been proved. In a quantum computation environment, most of these problems will be
no more difficult \cite{Shor94}, then the related public-key
protocols will not be secure.

Then we need to find new public-key schemes to resist the
attacks of quantum adversaries. One solution is private-key
protocols with the aid of quantum key distribution (QKD). Many
schemes of QKD have been proposed \cite{5,6,7,8,9}. Another
solution is to construct quantum public-key encryption (QPKE).
Okamoto et al. \cite{Okamoto00} introduce a public-key
encryption scheme with a quantum algorithm in key generation
phase. Gottesman \cite{Gottesman00} is the first to put forward
a protocol named ``quantum public key cryptography with
information-theoretic security''. Yang et al. \cite{Yang03, Liang10}
present some public-key encryption protocols based on
induced trap-door one-way transformation. Kawachi et al. present a
QPKE scheme of based on ``computational indistinguishability'' of two
quantum states \cite{Kawachi05,Kawachi04} related with automorphism
group of graphs problem. Nikolopoulos presents a QPKE scheme
\cite{Nikolopoulos08,Nikolopoulos09} based on single-qubit rotations.

This paper is organized as follows. In Section 2, we
describe some basic definitions of information-theoretic security, one-way function and hash functions.
In Section 3, we present two public-key encryptions orienting one-bit message based on conjugate coding. In section 4, we extend the schemes to those orienting multi-bits. In section 5, we describe a way to conceal the classical part of the public key with quantum state, this idea is expected to enhance a scheme to be information-theoretically secure. In section 6, we discuss a public-key encryption scheme based on a special entanglement state we presented in \cite{Pan10}.

\section{Preliminary}
\subsection{Information-theoretic security}
The ciphertext-indistinguishability under chosen plaintext
attack (IND-CPA) \cite{Menezes97} can be understand as that
while the adversary chooses two plaintexts and is then given
one of the corresponding ciphertexts, the adversary cannot yet
determine which plaintext corresponds to the previously unseen
ciphertext he received.

Strictly speaking, the concept information-theoretic IND-CPA is
defined as \cite{Goldreich04}: for every circuit family
$\{C_n\}$, every positive polynomial $p(\cdot)$, all
sufficiently large $n$, and every $x,y\in\{0,1\}^*$, the
probability $\operatorname*{Pr(\cdot)}$ satisfies:
 {\footnotesize\begin{equation}
|\operatorname*{Pr}[C_n(G(1^n),E_{G(1^n)}(x))=1]-\operatorname*{Pr}[C_n(G(1^n),E_{G(1^n)}(y))=1]|<\frac{1}{p(n)}.
\end{equation}}
In the case of QPKE, the information-theoretic quantum IND-CPA
is defined as \cite{Pan10}:
\begin{definition}
A quantum public-key encryption scheme is
information\\-theoretically ciphertext-indistinguishable under
quantum CPA if for every quantum circuit family $\{C_n\}$,
every positive polynomial $p(\cdot)$, all sufficiently large
$n$, and every bit-string $x,y\in\{0,1\}^*$, the probability
$\operatorname*{Pr(\cdot)}$ satisfies:
 {\footnotesize
\begin{equation}
\label{e4}
|\operatorname*{Pr}[C_n(G(1^n),E_{G(1^n)}(x))=1]-\operatorname*{Pr}[C_n(G(1^n),E_{G(1^n)}(y))=1]|<\frac{1}{p(n)},
\end{equation}}
where the algorithm $E$ is a quantum encryption algorithm, and
the ciphertext $E(x)$, $E(y)$ are quantum states.
\end{definition}

It is proved \cite{Yang12} that if the trace distance between
any two ciphertexts is $O(\frac{1}{2^n})$,
Eq.~(\ref{e4}) will hold, and the quantum public-key encryption
scheme will be information-theoretically secure.

The term ``bounded information-theoretically secure'' means
that the bit number of plaintext encrypted with
information-theoretic security has an upper bound. In
\cite{Kawachi04}, this bound of the protocol is proven to be
the bit number of private key \cite{Hayashi08}, thus it is much
less than that of a practical protocol.

\subsection{One-way function}
A trapdoor one-way function is necessary in a public-key
encryption protocol. A one-way function(OWF)based on classical
computational complexity hypothesis \cite{Menezes97} is a
function $f$ such that for each $x$ in the domain of $f$, it is
easy to compute $f(x)$; but for essentially all $y$ in the
range of $f$, it is infeasible to find any $x$ such that
$y=f(x)$ in expected polynomial time, i.e.
\begin{equation} \label{e2}
\operatorname*{Pr}[A(f(U_n),1^n)\in f^{-1}(f(U_n))]<\frac{1}{p(n)}.
\end{equation}
A trapdoor one-way
function \cite{Menezes97} is a one-way function $f$ with the
additional property that if some extra information (called
the trapdoor information) is given, it will becomes feasible to find an $x$
in the domain of $f$, for any given $y$ in the range of $f$,
such that $f(x)=y$.

The one-way property of classical OWF is usually in the sense
of computational security, i.e., it is based on a hypothesis of
the adversary's power of computation. The quantum one-way
transformation(OWT) we suggest here is in the sense of
information-theoretic security, since it is based on the
property of quantum states: people with the correct basis can
get the exact bits contained in the states, but one without the
information of basis cannot get them with an in-negligible probability.

\subsection{Hash functions and randomness}

In our schemes below, a multi-output Boolean function $F:\{0,1\}^m \rightarrow \{0,1\}^n (m>n)$ is used as the private key. Since $m>n$, $F$ can also be taken as a Hash function. Suppose $X$ is the set of all messages (input), and $Y$ is the set of all message digests (output) with $|Y|=M$. The theorem in \cite{Stinson03} ensures the randomness of the function's output:
\begin{theorem}
Suppose Hash function $h$ is randomly selected, let $X_0\subseteq X$. Suppose that if and only if $x\in X_0$, $h(x)$ is determined (through querying the oracle). Then for all $x\in X/X_0$ and $y\in Y$, there is always the relation $\operatorname*{Pr}[h(x)=y]=1/M$.
\end{theorem}

\section{Quantum public-key encryption schemes for one-bit message} \label{s1}
Let $k=(k_1,\cdots,k_n)$ be a n-bit string, where
$k_1,\cdots,k_n \in \{0,1\}$ . For the Hadamard transform
{\footnotesize
$H=\frac{1}{\sqrt{2}}\left[\begin{array}{ll}1&1\\1&-1
\end{array}\right]$}, we define $H_k=H^{k_1}\otimes \cdots
\otimes H^{k_n}$, where $H^0$ is the unit operator $I$, and
$H^1$ is $H$. Similarly for the Pauli operator {\footnotesize
$Y=\left[\begin{array}{ll}0&-\rm i\\\rm i&0
\end{array}\right]$}, we define
{\footnotesize$Y_k=Y^{k_1}\otimes \cdots \otimes Y^{k_n}$}. Let
$\Omega_b=\{i\in \{0,1\}^n|i_1\oplus\cdots\oplus i_n=b\}$, here $b=0,1$.

\subsection{First scheme} \label{s3}


In this scheme, the OWT is a transformation
mapping the classical message $b$ to an unknown state
$$\rho_b=\sum_{P(j)=b}Y_j\left[\sum_F
H_{F(s)}\left(\sum_{P(i)=0}p_{iF}|i\rangle\langle i|\right)H_{F(s)}\right]Y_j.$$ The
trapdoor information is the basis string $k=F(s)$ on which the
quantum states are encoded. The private key is a Boolean
function $F$. Using $s$, the classical part of the public-key,
the private-key owner can get the trapdoor information $k$ by
$k=F(s)$. The scheme is shown as follows:

\begin{bfseries}
\flushleft{[Key Generation]}
\end{bfseries} During this phase, Bob can do as follows:
\begin{enumerate}
\renewcommand{\labelenumi}{(G\theenumi)}
\item Select randomly a multi-output Boolean function
    $F:\{0,1\}^m \rightarrow \{0,1\}^n$ as his private key;
\item Select randomly $s\in \{0,1\}^m$, and computes
    $k=F(s)$;
\item Generate $|i\rangle$ with $i\in\Omega_0$;
\item Apply $H_k$ to $|i\rangle$, and take the
    classical-quantum pair $(s,H_k|i\rangle)$ as one of his
    public-keys.
\end{enumerate}

\begin{bfseries}
\flushleft{[Encryption]}
\end{bfseries}If Alice wants to send one bit
message $b$ to Bob, she should get one of Bob's public
keys, and then:
\begin{enumerate}
\renewcommand{\labelenumi}{(E\theenumi)}
\item Select $j$ randomly from
    $\Omega_b$;
\item Alice applies $Y_j$ to $H_k|i\rangle$, and then
    sends $(s,Y_j H_k|i\rangle)$ to Bob.
\end{enumerate}

\begin{bfseries}
\flushleft{[Decryption]}
\end{bfseries}
After receiving the ciphertext, Bob should:
\begin{enumerate}
\renewcommand{\labelenumi}{(D\theenumi)}
\item Calculate $k=F(s)$;
\item Apply $H_k$ to $Y_jH_k|i\rangle$, and measure it
    in the basis $\{|0\rangle,|1\rangle\}^n$.
\end{enumerate}

Here $F$ can be generated efficiently. Each output of the Boolean function $F$ can be written as
\begin{equation}
F^i(s)=\bigoplus_{d_1,\cdots,d_m}\alpha^i_{d_1,\cdots,d_m}s_1^{d_1}\cdots s_m^{d_m},
\end{equation}
where $i=1,\cdots,n$. There are $2^m$ terms to add up in each
$F^i(s)$, but after $m$ times of coin tossing, we can determined a $s_1^{d_1}\cdots s_m^{d_m}$ by
one instance of $d_1,\cdots,d_m$. We let the corresponding
$\alpha^i_{d_1,\cdots,d_m}=1$, then one component is determined. If
we toss the coin for $m\cdot O(m)$ times, $O(m)$ components are
determined. We then let other component coefficients to be zero. Thus $F$ can be
efficiently generated by $n\cdot m\cdot O(m)$. This algorithm
produces many strings of $m$ random bits, and the number of the
strings is polynomial in $m$.

Bob generates a large amount of public-keys with a private key
$F$. In each of these public-keys, $s$ is different. The
classical string $s$ is bounded to the quantum state
$H_k|i\rangle$ as a label.

The public key state for the adversary is
$
\sum_F\sigma_{F,s},
$
here \begin{eqnarray}\label{eq3}
\sigma_{F,s}=H_{F(s)}\left(\sum_{P(i)=0}p_{iF}|i\rangle\langle i|\right)H_{F(s)}.
\end{eqnarray}

The ciphertext state for the adversary is
\begin{equation}
\left(s,\sum_jY_j\left[\sum_F
H_{F(s)}\left(\sum_{P(i)=0}p_{ijF}|i\rangle\langle
i|\right)H_{F(s)}\right]Y_j\right),
\end{equation} and the quantum part of the ciphertext for a selected $j$ is
\begin{equation}
\rho_j=Y_j\left[\sum_F
H_{F(s)}\left(\sum_{P(i)=0}p_{iF}|i\rangle\langle
i|\right)H_{F(s)}\right]Y_j.
\end{equation}

While Bob gets the one-bit message from the parity of bit string he received. The state after applying $H_k$ to $Y_jH_k|i\rangle$ is
\begin{eqnarray}
H_kY_jH_k|i\rangle=(-1)^{k\cdot j}Y_j|i\rangle=(-1)^{j\cdot (i\oplus k)+ \frac{1}{2}W_H(j)}|i\oplus j\rangle.
\end{eqnarray}
Thus, after measuring it, Bob gets $|i\oplus j\rangle$, here $\oplus$ is bit-wise addition modulo 2. The parity of $W_H(j)$ is equal to the parity of $W_H(i\oplus j)$, because $W_H(i)$ is even. Then the message (plaintext) is obtained.

\subsection{Security analysis}
The adversary has two ways to attack the QPKE scheme. One is to
attack plaintext via distinguishing the two ciphertexts;
another is to attack the private key $F$ via getting
information of $k$.

For the first way of attack, we now prove that the trace
distance between the two different ciphertexts is
$O(\frac{1}{2^n})$.

The quantum part of public key
$H_k|i\rangle$ is a state consisting of $n$ qubits from the set
$\{|0\rangle,|1\rangle,|+\rangle,|-\rangle\}$, and the total
number of $|1\rangle$ and $|-\rangle$ is even. After encrypted
by Alice, the state is also $n$ qubits from the set
$\{|0\rangle,|1\rangle,|+\rangle,|-\rangle\}$.

If the message is 0, the number of 1 in $j$ is even, then the
number of qubits operated by $Y$ is even. If the number of $Y$
on $|1\rangle$ and $|-\rangle$ is odd, then the number of $Y$
on $|0\rangle$ and $|+\rangle$ (which producing $|1\rangle$ and
$|-\rangle$) is odd. Remember that the $|1\rangle$ and
$|-\rangle$ which are unchanged is odd, then after encrypted by
Alice, the total number of $|1\rangle$ and $|-\rangle$ will be
even. Given $s$ which is randomly selected, because of the way
$F$ is generated, the randomness of $k$ can be ensured. The
analysis when the message is 1 is similar. Then we can write
the states of ciphertext when the message is $b$ to be:
\begin{equation} \label{e1}
\rho_b=\frac{1}{2^{2n-1}}\sum_{{P(i)=b},j} |\psi_{i_1j_1}\rangle \langle \psi_{i_1j_1}| \cdots |\psi_{i_nj_n}\rangle \langle \psi_{i_nj_n}|,
\end{equation}
where $|\psi_{ij}\rangle\equiv H^j X^i|0\rangle$.

Now we calculate the trace distance between $\rho_0$ and
$\rho_1$ using the method in \cite{Yang10}. Define two
trace-preserving quantum operations $\mathcal{E}_1$ and
$\mathcal{E}_2$, for any n-bit quantum state $\rho$
\begin{equation}
\mathcal{E}_1(\rho)=U_{\frac{\pi}{4}}^{\otimes n}\rho {U_{\frac{\pi}{4}}^{\otimes n}}^\dagger,
\end{equation}
and
\begin{equation}
\mathcal{E}_2(\rho)=\frac{1}{2^n}\sum_{k\in\{0,1\}^n} H_k \rho H_k^\dagger.
\end{equation}
Here {\footnotesize $U_{\frac{\pi}{4}}=\rm e^{-\rm
i\frac{\pi}{4}Y}=\frac{\sqrt2}{2}\left[\begin{array}{ll}1&-1\\1&1
\end{array}\right]$}, which rotates each qubit of $\rho$ around
y axis by an angle of $\pi/2$:
\begin{eqnarray}
&U_{\frac{\pi}{4}}|0\rangle&\rightarrow~~|+\rangle,\nonumber\\
&U_{\frac{\pi}{4}}|1\rangle&\rightarrow-|-\rangle,\nonumber\\
&U_{\frac{\pi}{4}}|+\rangle&\rightarrow~~|1\rangle,\\
&U_{\frac{\pi}{4}}|-\rangle&\rightarrow~~|0\rangle.\nonumber
\end{eqnarray}

Define
\begin{equation}
\sigma_0=\frac{1}{2^{n-1}}\sum_{P(j)=0} |\phi_{j_1}\rangle \langle \phi_{j_1}| \cdots |\phi_{j_n}\rangle \langle \phi_{j_n}|,
\end{equation}
where $|\phi_j\rangle=H^j|0\rangle$.

We can see
\begin{eqnarray}
&&\mathcal{E}_2 \comp \mathcal{E}_1 (\sigma_0)\nonumber\\
=&&\mathcal{E}_2(U_{\frac{\pi}{4}}^{\otimes n}\sigma_0 {U_{\frac{\pi}{4}}^{\otimes n}}^\dagger)\nonumber\\
=&&\frac{1}{2^n}\sum_k H_k (U_{\frac{\pi}{4}}^{\otimes n}\sigma_0 {U_{\frac{\pi}{4}}^{\otimes n}}^\dagger) H_k^\dagger.
\end{eqnarray}
Because
\begin{numcases}
{H^iU_{\frac{\pi}{4}}|0\rangle=}|+\rangle~~(i=0)\nonumber\\
|0\rangle~~~(i=1)\nonumber
\end{numcases}
\begin{numcases}
{H^iU_{\frac{\pi}{4}}|+\rangle=}|1\rangle~~~(i=0)\nonumber\\
|-\rangle~~(i=1)\nonumber
\end{numcases}
then
\begin{eqnarray}
&&\mathcal{E}_2 \comp \mathcal{E}_1 (\sigma_0)\nonumber\\=&&\frac{1}{2^{2n-1}}\sum_{P(i)=0, j} |\psi_{i_1j_1}\rangle \langle \psi_{i_1j_1}| \cdots |\psi_{i_nj_n}\rangle \langle \psi_{i_nj_n}|.
\end{eqnarray}
That means
\begin{equation}
\mathcal{E}_2 \comp \mathcal{E}_1 (\sigma_0)=\rho_0.
\end{equation}
And similarly we define
\begin{equation}
\sigma_1=\frac{1}{2^{n-1}}\sum_{P(j)=1} |\phi_{j_1}\rangle \langle \phi_{j_1}| \cdots |\phi_{j_n}\rangle \langle \phi_{j_n}|.
\end{equation}
We can get
\begin{equation}
\mathcal{E}_2 \comp \mathcal{E}_1 (\sigma_1)=\rho_1.
\end{equation}
As trace-preserving quantum operations are contractive
\cite{Nilsen00}, we get
\begin{equation}
D(\rho_0,\rho_1) \le D(\sigma_0,\sigma_1).
\end{equation}

$D(\sigma_0,\sigma_1)$ is easy to compute. By Mathematical
induction, we have
\begin{eqnarray}
\sigma_0-\sigma_1&=&\frac{1}{2^{n-1}}(|0\rangle\langle0|-|+\rangle\langle+|)^{\otimes n}\nonumber\\
&=&\frac{1}{2^{n-1}}\left[\begin{array}{cc}
\frac{1}{2}&-\frac{1}{2}\\
-\frac{1}{2}&-\frac{1}{2}
\end{array}\right]^{\otimes n}.
\end{eqnarray}
Then
\begin{eqnarray}
D(\sigma_0,\sigma_1)&=&\frac{1}{2}\cdot \rm tr \left|\sigma_0-\sigma_1\right|\nonumber\\
&=&\frac{1}{2^{n}}\rm tr\left|\left[\begin{array}{cc}
\frac{1}{2}&-\frac{1}{2}\\
-\frac{1}{2}&-\frac{1}{2}
\end{array}\right]^{\otimes n}\right|\nonumber,
\end{eqnarray}
here $\left|A\right|$ is the singular value matrix of matrix
$A$, $\left|A\right|=\sqrt{A^\dagger A}$. By spectral
decomposition, we have the following conclusion for normal
matrices $A_1$ and $A_2$:
\begin{equation}
\rm tr|A_1\otimes A_2|=\rm tr|A_1|\cdot \rm tr|A_2|,
\end{equation}
we then have
\begin{eqnarray}
D(\sigma_0,\sigma_1)&=&\frac{1}{2^{n}}\left(\rm tr\left|\left[\begin{array}{cc}
\frac{1}{2}&-\frac{1}{2}\\
-\frac{1}{2}&-\frac{1}{2}
\end{array}\right]\right|\right)^n\nonumber\\
&=&(\frac{\sqrt{2}}{2})^n.
\end{eqnarray}
Finally
\begin{equation}
D(\rho_0,\rho_1) \le D(\sigma_0,\sigma_1)=(\frac{\sqrt{2}}{2})^n.
\end{equation}

In order to against the second kind of attack, the scheme can be amended as follow.

\subsection{Second scheme}\label{s0}
In the first scheme, half of $i$ is impossible to appear, here we show a enhance scheme that ensures $i$'s randomness.

\begin{bfseries}
\flushleft{[Key generation]}
\end{bfseries}During key generation phase, Bob
generates his private and public keys, he can do as follows:
\begin{enumerate}
\renewcommand{\labelenumi}{(G\theenumi)}
\item Generate randomly a multi-output Boolean function $F=(F_1, F_2)$ as his
    private key, here $F_1:\{0,1\}^m \rightarrow \{0,1\}^n$ and $F_2:\{0,1\}^m \rightarrow \{0,1\}$ is a balanced Boolean
    function;
\item Select randomly $|i\rangle\in \{0,1\}^n$. Then select
    $s\in \{0,1\}^m$ randomly, and compute $k=F_1(s),p=F_2(s)$.
    If $p\neq P(i)$, Bob selects $s$
    again until $p=P(i)$;
\item Apply $H_k$ to $|i\rangle$, and take $(s,
    H_k|i\rangle)$ as one of his public-key.
\end{enumerate}
\begin{bfseries}
\flushleft{[Encryption]}
\end{bfseries}If Alice wants to send one bit
message $b$ to Bob, she should get one of Bob's public keys,
and then:
\begin{enumerate}
\renewcommand{\labelenumi}{(E\theenumi)}
\item Select $j$ randomly from
    $\Omega_b$;
\item Alice applies $Y_j$ to $H_k|i\rangle$, and then
    sends $(s,Y_j H_k|i\rangle)$ to Bob.
\end{enumerate}

\begin{bfseries}
\flushleft{[Decryption]}
\end{bfseries}After Bob receives
$(s,Y_j H_k|i\rangle)$ sent by Alice, he should:
\begin{enumerate}
\renewcommand{\labelenumi}{(D\theenumi)}
\item Calculate $(k, P(i))=F(s)$;
\item Apply $H_k$ to $Y_jH_k|i\rangle$, and measures it in
    the basis $\{|0\rangle,|1\rangle\}^n$.
\end{enumerate}

$F$ can be generated efficiently by the similar local coin tossing
algorithm in \ref{s3}. In addition, to ensure the
randomness of $i$, the set $\mathcal{F}$ made of $F_2$ should satisfy that: for any $f\in\mathcal{F}$,
$f+1\in\mathcal{F}$. To achieve this property, we only need to add one extra coin tossing for determining whether add an extra ``1" to $F_2$.
Furthermore, for a fixed $s$, the probability of $i$ distributes evenly to the adversary.

Let $\operatorname*{P}(i)=P(i)$ is the parity bit of $i$.
The state after applying $H_k$ to $Y_jH_k|i\rangle$ is $(-1)^{j\cdot (i\oplus k)+W_H(j)/2}|i\oplus j\rangle.$
Thus, after measuring $H_kY_jH_k|i\rangle$, Bob gets $|i\oplus
j\rangle$. If $\operatorname*{P}(i)=0$, the parity of $W_H(j)$
is equal to the parity of $W_H(i\oplus j)$; if
$\operatorname*{P}(i)=1$, the parity of $W_H(j)$ is opposite to
the parity of $W_H(i\oplus j)$. Then the message (0 or 1) is
obtained from the parity of $W_H(j)$.

\subsection{Security analysis} \label{s2}
For the attack to plaintext, the adversary cannot get any
information of it from the ciphertext. Let the
ciphertexts for a fixed $s$ and $b$ is $\rho_b(s)$, first we divide $\mathcal{F}$ into two part: $\mathcal{F}_s^0$ and $\mathcal{F}_s^1$, they satisfy that $$\mathcal{F}_s^0=\{f\in\mathcal{F}|f(s)=0\},$$ and
\begin{eqnarray}
  \mathcal{F}_s^1=\{f\in\mathcal{F}|f(s)=1\}.
\end{eqnarray}
Then we get $|\mathcal{F}_s^0|=|\mathcal{F}_s^1|=\frac{|\mathcal{F}|}{2}$, and
\begin{equation}
\rho_b=\sum_{P(j)=b}p_{j}Y_j\left[\sum_{F_1,F_2}\sum_ip_{F_1}p_ip_{F_2|i}H_{F_1(s)}|i\rangle\langle i|H_{F_1(s)}\right]Y_j.
\end{equation}
Here $p_j=\frac{1}{2^{n-1}}$, $p_{F_1}=\frac{1}{|\{F_1\}|}$, $p_i=\frac{1}{2^n}$ and $p_{F_2|i}$ is conditional probability of $F_2$ while $i$ and $s$ is fixed,
\begin{eqnarray}
  p_{F_2|i}=\left\{\begin{array}{cc}
                 0 & (F_2\in\mathcal{F}_s^b) \\
                 \frac{2}{|\mathcal{F}|} & (F_2\in\mathcal{F}_s^{\bar{b}}).
               \end{array}\right.
\end{eqnarray}
Thus we have: $\sum_{F_2}p_{F_2|i}=1$ and
\begin{equation}
\rho_b=\sum_{P(j)=b}p_{j}Y_j\sum_{F_1}\sum_i\left[p_{F_1}p_iH_{F_1(s)}|i\rangle\langle i|H_{F_1(s)}\sum_{F_2}p_{F_2|i}\right]Y_j=\frac{I}{2^n},
\end{equation}
so
\begin{eqnarray}\label{eq4}
D(\rho_0,\rho_1)=0.
\end{eqnarray}

Denote $$\rho_{F,s,b}=\sum_{P(j)=b}\sum_ip_{j}p_{i|F_2}Y_j\left[H_{F_1(s)}\left(|i\rangle\langle i|\right)H_{F_1(s)}\right]Y_j,$$
\begin{eqnarray}
  p_{i|F_2}=\left\{\begin{array}{cc}
                 0 & (P(i)\neq F_2(s)) \\
                 \frac{1}{2^{n-1}} & (P(i)=F_2(s)).
               \end{array}\right.
\end{eqnarray}
when this scheme is used to encrypt $t$ bits, the ciphertext for the bit-string can be written as
\begin{equation}
\sum_F\left(\rho_{F,s_1,b_1}\otimes\cdots\otimes \rho_{F,s_t,b_t}\right).
\end{equation}
Thus the security is depend on this trace distance:
\begin{equation}
D\left(\sum_F\left(\rho_{F,s_1,b_1}\otimes\cdots\otimes \rho_{F,s_t,b_t}\right),\sum_F\left(\rho_{F,s_1,b'_1}\otimes\cdots\otimes \rho_{F,s_t,b'_t}\right)\right).
\end{equation}

Even the Eq.~\ref{eq4} hold, this does not mean the scheme is information-theoretic secure for one-bit oriented encryption.
In practical attack, the adversary may own some public keys as extra information, the density operator of public key is:
$\sum_F\tau_{F,s}$, here
\begin{eqnarray}
  \tau_{F,s}=H_{F_1(s)}\sum_{i}(p_{i|F_2}|i\rangle\langle i|)H_{F_1(s)}.
\end{eqnarray}
so the trace distance for different one-bit ciphertexts is:
\begin{eqnarray}\label{eq5}
 D\left(\sum_F(\rho_{F,s_0,0}\otimes\tau_{F,s_1}\otimes\cdots\otimes\tau_{F,s_t}),
  \sum_F(\rho_{F,s_0,1}\otimes\tau_{F,s_1}\otimes\cdots\otimes\tau_{F,s_t}) \right).
\end{eqnarray}
The trace distance for different m-bits string ciphertexts should be
\begin{eqnarray}\label{eq6}
 D\left(\sum_F(\rho_{F,\vec{s},\vec{b}}\otimes\tau_{F,\vec{s'}}),\sum_F(\rho_{F,\vec{s},\vec{b'}}\otimes\tau_{F,\vec{s'}})\right),
\end{eqnarray}
where
\begin{eqnarray}
\rho_{F,\vec{s},\vec{b}}=\rho_{F,s_1,b_1}\otimes\cdots\otimes \rho_{F,s_t,b_t},
\end{eqnarray}
\begin{eqnarray}
\tau_{F,\vec{s'}}=\tau_{F,s'_1}\otimes\cdots\otimes\tau_{F,s'_t}.
\end{eqnarray}

Thus if we want to prove the scheme is information-theoretic secure, it follows that Eqs.~(\ref{eq5})(\ref{eq6}) are exponential small according to \cite{Yang12}.

\subsection{Discussion on quantum OWT}

To do the OWT from message $b$ to state
$\mathcal{E}_b(\rho_0^{(n)}(F(s),i))$, one usually does unitary
operation depending on $b$ on the public key {\footnotesize
$\rho_0^{(n)}(F(s),i)$}, then discard the extra outputs and get
$\mathcal{E}_m(\rho_0^{(n)}(F(s),i))$. One cannot obtain the
original state without the extra outputs. If one has done many
operations on $\rho_0^{(n)}(F(s),i)$, he cannot get the
original state even with the extra outputs, because he does not
know the corresponding state of extra outputs for the certain
operation.

Consider the unitary transformation
$U_f(|x\rangle|y\rangle)=|x\rangle|y\oplus f(x) \rangle$.
$|x\rangle$ is the input qubit, and $|y\rangle$ is auxiliary
qubit. For the initial state $\sum_{x_i}
\alpha_{x_i}|x_i\rangle |0\rangle$, the output is
\begin{equation}
U_f(\sum_{x_i}
\alpha_{x_i}|x_i\rangle |0\rangle)=\sum_{x_i}
\alpha_{x_i}|x_i\rangle |f(x_i)\rangle.
\end{equation}
People with trapdoor information $|x_i\rangle$ can do the
following operation
\begin{equation}
U_{f^{-1}}|x_i\rangle|f(x_i)\rangle=|0\rangle|f(x_i)\rangle.
\end{equation}
For people without trapdoor information $|x_i\rangle$, he can
only do as
\begin{equation}
U_f(|x_i\rangle|f(x_j)\rangle)=|x_i\rangle|f(x_j)\oplus f(x_i)
\rangle.
\end{equation}
If $d_H(f(x_i),f(x_j))=0$, $|f(x_j)\oplus f(x_i)
\rangle=|0\rangle$, then the input for $f(x_j)$ is obtained, here $d_H$ is the Hamming distance.
Consider the probability when $d_H(f(x_i),f(x_j))=0$ is valid.
For a given $f(x_j)$, the probability is $(\frac{1}{2})^n$.
Then according to Eq.~(\ref{e2}), it is a one-way
transformation.

\section{Extend to multi-bit-oriented schemes}
The scheme in Sec.~\ref{s1} is one-bit-oriented. We now extend
it to multi-bit-oriented schemes. In the following scheme of
multi-bit message, the OWT is to map the classical message $j$
to an unknown state$$\rho_{j}=Y_j\left[\sum_F
H_{F(s)}\left(\sum_ip_{iF}|i\rangle\langle
i|\right)H_{F(s)}\right]Y_j.$$ The trapdoor information is the
basis $k=F(s)$ on which the quantum states are encoded. The
private key is a Boolean function $F$. Using a part of the
public-key $s$, the owner of private-key can get the trapdoor
information $k$ by $k=F(s)$.

\subsection{First scheme}
\begin{bfseries}
\flushleft{[Key generation]}
\end{bfseries} During this phase, Bob can do as follows:
\begin{enumerate}
\renewcommand{\labelenumi}{(G\theenumi)}
\item Select randomly a multi-output Boolean function $F:\{0,1\}^m \rightarrow
    \{0,1\}^n$ as his private key;
\item Select randomly $s_1,s_2\in \{0,1\}^m$, and compute $k=F(s_1)$,
    $i=F(s_2)$;
\item Apply $H_k$ to $|i\rangle$, and take
    $(s_1, s_2,H_k|i\rangle)$ as his public key.
\end{enumerate}

\begin{bfseries}
\flushleft{[Encryption]}
\end{bfseries}
If Alice wants to send n-bit message $j$ to Bob, she should get one of Bob's public keys, and then:
\begin{enumerate}
\renewcommand{\labelenumi}{(E\theenumi)}
\item Apply $Y_j$ to $H_k|i\rangle$, and then sends $(s_1, s_2,Y_jH_k|i\rangle)$ to Bob.
\end{enumerate}

\begin{bfseries}
\flushleft{[Decryption]}
\end{bfseries}
After Bob receives $|\psi_{s_1}\rangle\otimes
|\psi_{s_2}\rangle\otimes Y_jH_k|i\rangle$ sent by Alice, he
should:
\begin{enumerate}
\renewcommand{\labelenumi}{(D\theenumi)}
\item Calculate $k=F(s_1)$, $i=F(s_2)$;
\item Apply $H_k$ to $Y_jH_k|i\rangle$, and measure
    on the basis $\{|0\rangle,|1\rangle\}^n$.
\end{enumerate}

\subsection{Second scheme} \label{s5}

\begin{bfseries}
\flushleft{[Key generation]}
\end{bfseries}
\begin{enumerate}
\renewcommand{\labelenumi}{(G\theenumi)}
\item Select randomly two multi-output Boolean function $F_1:\{0,1\}^n \rightarrow
    \{0,1\}^n$,
$F_2:\{0,1\}^n \rightarrow \{0,1\}^n$ as his private key;
\item Select randomly $s\in \{0,1\}^m$, and compute
    $k=F_1(s)$, $i=F_2(s)$;
\item Apply $H_k$ to $|i\rangle$, and and take
    $(s,H_k|i\rangle)$ as his one
    public-key.
\end{enumerate}

\begin{bfseries}
\flushleft{[Encryption]}
\end{bfseries}
If Alice wants to send n-bit message $j$ to Bob, she should get one of Bob's public keys, and then:
\begin{enumerate}
\renewcommand{\labelenumi}{(E\theenumi)}
\item Apply $Y_j$ to $H_k|i\rangle$, and then sends $(s,Y_jH_k|i\rangle)$ to Bob.
\end{enumerate}

\begin{bfseries}
\flushleft{[Decryption]}
\end{bfseries}
After Bob receives $(s,Y_jH_k|i\rangle)$ sent by Alice, he should:
\begin{enumerate}
\renewcommand{\labelenumi}{(D\theenumi)}
\item Calculate $k=F_1(s)$, $i=F_2(s)$;
\item Apply $H_k$ to $Y_jH_k|i\rangle$, and measure
    on the basis $\{|0\rangle,|1\rangle\}^n$.
\end{enumerate}

\subsection{Security analysis for both schemes}
In the two multi-bit-oriented schemes described above, Bob can
get the message via measuring the result. The state after
applying $H_k$ to $Y_jH_k|i\rangle$ is $(-1)^{j\cdot (i\oplus
k)+W_H(j)/2}|i\oplus j\rangle$, thus, after measuring
$H_kY_jH_k|i\rangle$, Bob gets $|i\oplus j\rangle$. Because Bob
can get the exact value of $i$, he can get the message $j$
finally.

The adversary has also two ways to attack these two schemes.
One is to attack the private key via getting information of
$k$; another is to attack plaintext via distinguishing the two
ciphertexts. For the first method, similar to the analysis in
Sec.~\ref{s2}, $H_k|i\rangle$ for $k$ and $k'$ $(k\neq k')$ are
$\rho_k=\frac{1}{2^n}\sum_{i\in\{0,1\}^n}H_k|i\rangle\langle i|
H_k^{\dagger}$ and
$\rho_{k'}=\frac{1}{2^n}\sum_{i\in\{0,1\}^n}H_{k'}|i\rangle\langle
i| H_{k'}^{\dagger}$, respectively. We can get
$\rho_k=\rho_{k'}=I/2^n$, then
$D(\mathcal{E}(\rho_k),\mathcal{E}(\rho_{k'}))\leq
D(\rho_k,\rho_{k'})=0$. Thus no quantum algorithm can
distinguish the two quantum states $\rho_k$ and $\rho_{k'}$,
the two states are indistinguishable.

For the second way, we can prove that the adversary cannot get
any information of the plaintext from the ciphertext. Let the
message be $j$, the ciphertext is
\begin{equation}
\rho_{j}=Y_j\left[\sum_F
H_{F(s)}\left(\sum_ip_{iF}|i\rangle\langle
i|\right)H_{F(s)}\right]Y_j=I/2^n.
\end{equation}
Then for any $j$ and $j'$$(j\neq j')$, $D(\rho_j,\rho_{j'})=0$.
Thus $D(\mathcal{E}(\rho_j),\mathcal{E}(\rho_{j'}))\leq
D(\rho_j,\rho_{j'})=0$. No quantum algorithm can distinguish
the two ciphertext. According to \cite{Yang12}, this scheme is
information-theoretic secure.

\section{Enhanced scheme} \label{s4}
In the schemes given above, though $H_k|i\rangle$ is unknown to the adversary, he can point out whether each private key is different since $s$ is obvious, so it generally requires that the same public-key should not be reused. Here we show a method to solve this problem, enhance the scheme in
Sec.~\ref{s0} for example:
\begin{bfseries}
\flushleft{[Key generation]}
\end{bfseries}During key generation phase, Bob
generates his private and public keys, he can do as follows:
\begin{enumerate}
\renewcommand{\labelenumi}{(G\theenumi)}
\item Generate a multi-output Boolean function $F:\{0,1\}^m
    \rightarrow \{0,1\}^{n+1}$ randomly, then select randomly $l\in \{0,1\}^m$. Take $(F,l)$ as his private key;
\item Select randomly $|i\rangle\in \{0,1\}^n$. Then select
    $s\in \{0,1\}^m$ randomly, and compute $(k, p)=F(s)$.
    If $p\neq P(i)$, Bob selects $s$
    again until $p= P(i)$;
\item Calculate
    $|\psi_s\rangle=|s_1\rangle_{l_1}\otimes|s_2\rangle_{l_2}\otimes\cdots
    \otimes|s_m\rangle_{l_m}$;
\item Apply $H_k$ to $|i\rangle$, and take
    $|\psi_s\rangle\otimes H_k|i\rangle$ as his one
    public-key.
\end{enumerate}


\begin{bfseries}
\flushleft{[Encryption]}
\end{bfseries}If Alice wants to send one bit
message $b$ to Bob, she should get one of Bob's public keys,
and then:
\begin{enumerate}
\renewcommand{\labelenumi}{(E\theenumi)}
\item Select $j$ randomly from
    $\Omega_b$;
\item Alice applies $I\otimes Y_j$ to
    $|\psi_s\rangle\otimes H_k|i\rangle$, and then sends
    $|\psi_s\rangle\otimes Y_j H_k|i\rangle$ to Bob.
\end{enumerate}

\begin{bfseries}
\flushleft{[Decryption]}
\end{bfseries}After Bob receives
$|\psi_s\rangle\otimes Y_j H_k|i\rangle$ sent by Alice, he
should:
\begin{enumerate}
\renewcommand{\labelenumi}{(D\theenumi)}
\item Calculate from $|\psi_s\rangle$ and $l$ to get $s$;
\item Calculate $(k, P(i))=F(s)$;
\item Apply $H_k$ to $Y_jH_k|i\rangle$, and measures it in
    the basis $\{|0\rangle,|1\rangle\}^n$.
\end{enumerate}

This method is also useful for our other schemes. It can be seen that, if we use $(s, H_k|i\rangle)$ as
public key, for the same $s$, $k$ is also the same for one-bit-oriented schemes, and $i$ is also the same for multi-bit-oriented schemes.
Then there may exist many copies of the same $(s, H_k|i\rangle)$, the adversary may obtain information of $F$
with these copies. While we use $|\psi_s\rangle\otimes H_k|i\rangle$ as the public key, the
adversary cannot get $s$ directly, then he cannot get information of $F$ even as $s$ are reused.

\section{Public-key encryption scheme based on a special entanglement state\cite{Pan10}}
In \cite{Pan10} we present a public-key encryption scheme based on a special entanglement state, we prove that the scheme is secure under the attack to plaintext, since the ciphertexts encrypted from different plaintexts are indistinguishable. Here we first complete the proof for Theorem 6 in \cite{Pan10}, which is
related to the attack to plaintext with ciphertext. Then we will give a way to attack the private key by the public key.

Let $\Omega_n=\{k\in Z_{2^n}|W_H(k)~is~odd\}$ and $\Pi_n=\{k\in Z_{2^n}|W_H(k)~is~even\}$, where $W_H(k)$ is $k$'s Hamming weight. Let a Boolean Function $F: \Omega_n\rightarrow\Omega_n$ be the private key. For a randomly chosen $s$, $(s, \rho_{k,i}^0)$ is the public key satisfies that $k=F(s)$ and
\begin{eqnarray}
  \rho_{k,i}^0=\frac{1}{2}(|i\rangle+|i\oplus k\rangle)(\langle i|+\langle i\oplus k|),
\end{eqnarray}
as the same as that of \cite{Pan10}.

Then the density operator of ciphertext for $b=0$ is $\frac{1}{2^{n-1}}\sum_k\rho_k^0$ for the adversary:
 \begin{eqnarray}
  \rho_{k}^0=\frac{1}{2\cdot 2^n}\sum_i2(|i\rangle\langle i|+|i\rangle\langle i\oplus k|)=\frac{1}{2^n}\sum_i\sum_x|i\rangle\langle i\oplus xk|,
\end{eqnarray}
where $x\in\{0,1\}$.

In order to analysis the attack to the plaintext, we should calculate the trace distance between ciphertexts encrypted from different plaintexts: $$\parallel\frac{1}{2^{n-1}}\sum_k((\rho_k^0-\rho_k^1)\otimes(\rho_k^0)^{\otimes t-1})\parallel_{\rm tr},$$
here the first part of the summation $(\rho_k^0-\rho_k^1)$ represent the difference between ciphertexts, and the latter part $(\rho_k^0)^{\otimes t}$ represent the extra $t-1$ public keys that the adversary would own.

First we give the the trace distance between $\frac{1}{2^{n-1}}\sum_k(\rho_k^0)^{\otimes t}$ and $(\frac{I}{2^n})^\otimes t$:
\begin{eqnarray}\label{eq1}
&&\parallel\frac{1}{2^{n-1}}\sum_k(\rho_k^0)^{\otimes t}-(\frac{I}{2^n})^\otimes t\parallel_{\rm tr}\nonumber\\
&=&\frac{1}{2^{n-1}2^{nt}}\parallel\sum_k\sum_i(\sum_{x}|i\rangle\langle i\oplus xk|-|i\rangle\langle i|)\parallel_{\rm tr}\nonumber\\
&=&\frac{1}{2^{n-1}2^{nt}}\parallel\sum_k\sum_i\sum_{x\neq(0\cdots0)}|i\rangle\langle i\oplus xk|\parallel_{\rm tr}.
\end{eqnarray}

Let $A=\sum_k\sum_i\sum_{x\neq(0\cdots0)}|i\rangle\langle i\oplus xk|$, it has a polar decomposition $A=|A|V$, then
$|{\rm tr}(AV^{\dag})|=|{\rm tr}|A||={\rm tr}|A|$, so the eq.~(\ref{eq1}) is equal to:
\begin{eqnarray}\label{eq2}
  &&\frac{1}{2^n2^{nt}}|tr(\sum_k\sum_i\sum_{x\neq(0\cdots0)}|i\rangle\langle i\oplus xk|V^{\dag})|\nonumber\\
  &=&\frac{1}{2^n2^{nt}}|\sum_i(\sum_k\sum_{x\neq(0\cdots0)}(\langle i\oplus xk|)V^{\dag}|i\rangle)|\nonumber\\
  &\leq&\frac{1}{2^n2^{nt}}\sum_i(\parallel V^{\dag}|i\rangle\parallel\cdot\parallel\sum_k\sum_{x\neq(0\cdots0)}\langle i\oplus xk|\parallel)\nonumber\\
  &\leq&\frac{1}{2^n2^{nt}}\cdot2^{nt}\cdot\sqrt{2^{n-1}(2^t-1)}<\sqrt{\frac{1}{2^{n-t+1}}}.
\end{eqnarray}

Similarly, we can get
\begin{eqnarray}
\parallel\frac{1}{2^{n-1}}\sum_k(\rho_k^1\otimes(\rho_k^0)^{\otimes t-1}-(\frac{I}{2^n})^{\otimes t})\parallel_{\rm tr}<\sqrt{\frac{1}{2^{n-t+1}}}.
\end{eqnarray}
Then we have
\begin{eqnarray}
& &\parallel\frac{1}{2^{n-1}}\sum_k((\rho_k^0-\rho_k^1)\otimes(\rho_k^0)^{\otimes t-1})\parallel_{\rm tr}\nonumber\\
&\leq&
\parallel\frac{1}{2^{n-1}}\sum_k(\rho_k^1\otimes(\rho_k^0)^{\otimes t-1}-(\frac{I}{2^n})^{\otimes t})\parallel_{\rm tr}
+\parallel\frac{1}{2^{n-1}}\sum_k(\rho_k^0)^{\otimes t}-(\frac{I}{2^n})^\otimes t\parallel_{\rm tr}\nonumber\\
&=&\sqrt{\frac{1}{2^{n-t-1}}}.
\end{eqnarray}
Thus the proof complete.\\

We will show a way to attack the private key by the public key. While the attacker gets a public key $(s, \rho_{k,i}^0)$, he can attack by taking a unitary operation as follow:
\begin{eqnarray}
  H^n\frac{1}{\sqrt{2}}(|i\rangle+|i\oplus k\rangle)&=&\frac{1}{\sqrt{2^{n+1}}}\sum_y\left((-1)^{y\cdot i}|y\rangle+(-1)^{y\cdot (i\oplus k)}|y\rangle\right)\nonumber\\
  &=&\frac{1}{\sqrt{2^{n+1}}}\sum_y(-1)^{y\cdot i}(1+(-1)^{y\cdot k})|y\rangle.
\end{eqnarray}

Then he measures the last state, if he gets a result called $y_0$, it must satisfy $y_0\cdot k=0$, thus he get a equation of $F$ that:
\begin{eqnarray}
F(S)\cdot y_0=0,
\end{eqnarray}
which contains on bit information of $F$.

$F$ is also built by $n^2\cdot p(n)$ variables:
$$F(s_1,\cdots, s_n)=(F^1(s),\cdots,F^n(s)),$$
where
$$F^j(s)=(s_1^{a_{j11}}\cdots s_n^{a_{j1n}})\oplus(s_1^{a_{j21}}\cdots s_n^{a_{j2n}})\oplus\cdots\oplus (s_1^{a_{jp(n)1}}\cdots s_n^{a_{jp(n)n}}),$$ each $a_{jkl}=G(1)$ is a result of a toss.
Since $x^a=xa\oplus a\oplus1$, There is a linear expression of $F$:
\begin{eqnarray}
F^j(s)=\bigoplus_{\alpha=1}^{p(n)}\left(\prod_{\beta=1}^n(s_{\beta}a_{j\alpha\beta}+a_{j\alpha\beta}+1)\right).
\end{eqnarray}
The scheme should be only called bounded information secure similar to the scheme in \cite{Okamoto00,Hayashi08}.

According to the schemes based on conjugate coding, it is secure under the attack to the private key, since that the attack to the private key is equivalent to the attack to the basis of an unknown state, we have already proved that if this attack can be achieved it will lead to a Superluminal communication\cite{Yang12}.

\section{Discussions}
Every classical public-key scheme, such as RSA \cite{Rivest78},
is insecure under man-in-the-middle (MIM) attack. If the
adversary can intercept Bob's public key distribution channel,
she can replace Bob's public-key $(n,e)$ with her own public
key $(n',e')$, and send it to Alice. While Alice encrypts her
message with $(n',e')$ and sends the ciphertext back to Bob,
the adversary may decrypt the ciphertext to get the message.
Finally the adversary encrypts the message with $(n,e)$, and
sends this ciphertext to Bob. We can see that nobody will be
aware of the exist of the adversary. It is clear that for the
design of a public-key encryption scheme, it is necessary to
provide that Alice can obtain the public-key of Bob securely.
Actually, this precondition is necessary for all classical and
quautum public-key encryption protocols. To resist the MIM
attack is the task of, such as public-key infrastructure (PKI).

A common feature of the scheme in \cite{Kawachi05} and our
bit-oriented scheme is: an n-qubit public key is needed to
encrypt a one-bit message. The difference is, in the scheme of
\cite{Kawachi05}, for different messages, the public keys are
the same; however, in our schemes, the public keys are
different for every time of encryption, as a result of choosing
different $s$ and $|i\rangle$ at each time of key generation.
The security of our schemes is highly improved, while we can
see that the resources needed to store the quantum public keys
are kept the same.

In symmetric cryptography, one-time pad is information-theoretically secure. In asymmetric cryptography, the schemes presented here can reach bounded information-theoretic security, and the different keys used at different time are public keys rather than private keys. However, in classical asymmetric cryptography, even if we use different public keys for every time of encryption, information-theoretic security still cannot be reached. For example, in the RSA scheme, if a different public key is used every new time of encryption, the private key changes accordingly. Then only two ciphertexts are needed to get the plaintext, and it is not secure.

\section{Conclusion}

We propose several QPKE schemes of classical messages based on conjugate coding, and
prove that some of them are bounded information-theoretic secure. We also discuss a scheme we presented in \cite{Pan10} based on the state $|i\rangle+|i\oplus k\rangle$ and give a way to attack it. These schemes include scheme present in \cite{Okamoto00} are all bounded information-theoretic secure, but the schemes based on conjugate coding seem to be more secure than those in \cite{Okamoto00,Pan10} as their public key's reuse times is limit to the number of bits of the private key. From a practical point of view, our schemes based on single-photon
string may be realized in near future.

\section*{Acknowledgement}

This work was supported by the National Natural Science
Foundation of China under Grant No. 61173157.













\end{document}